# Graphene Based Terahertz Phase Modulators

N Kakenov[1], M S Ergoktas[3], O Balci[1] and C Kocabas [1,3] [†]

[1] Department of Physics, Bilkent University, Ankara, 06800, Turkey

[2] Institute of Materials Science and Nanotechnology, Bilkent University, Ankara 06800, Turkey

[3] School of Materials and National Graphene Institute, University of Manchester, Oxford Rd, Manchester, M13 9PL, UK

[†]Email: coskun.kocabas@manchester.ac.uk

**Abstract.** Electrical control of amplitude and phase of terahertz radiation (THz) is the key technological challenge for high resolution and noninvasive THz imaging. The lack of an active materials and devices hinders the realization of these imaging systems. Here, we demonstrate an efficient terahertz phase and amplitude modulation using electrically tunable graphene devices. Our device structure consists of electrolyte-gated graphene placed at quarter wavelength distance from a reflecting metallic surface. In this geometry, graphene operates as a tunable impedance surface which yields electrically controlled reflection phase. Terahertz time domain reflection spectroscopy reveals the voltage controlled phase modulation of $\pi$ and the reflection modulation of 50 dB. To show the promises of our approach, we demonstrate a multipixel phase modulator array which operates as a gradient impedance surface.

## 1. Introduction

Electromagnetic waves (EMW) are described by vector fields which include polarization degree of freedom and spatial distribution of intensity and phase. Dynamic control of these physical quantities provides means of encoding information. Imaging and communication systems rely on the manipulation of light with electrical signals that can either directly control the generation of light at the source or it can indirectly manipulate the light during the propagation through a medium. For visible and infrared light, electrically controlled sources and rich variety of electro-optical controls have already been developed [1, 2]. These techniques enabled the realization of sophisticated optical devices. For long-wavelength light, radio waves and microwaves, the manipulation of the EMW during the propagation is a challenge. For terahertz radiation, this challenge is even more severe due to the requirement of ultrafast electronics and the lack of an active materials. Over the past two decades, a significant amount of research have been devoted to develop active devices which control intensity and phase of terahertz radiation [3-9]. Early studies are based on cooled-semiconductor quantum wells whose intersubband absorption falls into terahertz energies [10, 11]. Depletion of electrons from the quantum wells leads to the reduced absorption and enhanced reflection. Electromagnetic phenomena achieved by metamaterials enhances these interactions. Chen et al. demonstrated metamaterial based intensity and phase modulators [12]. Furthermore, these approaches were used to realize THz imaging systems [13].

Recent developments in the field of 2-dimentional crystals provide alternative solutions to control light-matter interaction in a very broad spectrum. Since light-matter interaction is mediated by high mobility electrons, electrostatic tuning of charge density on atomically thin layers facilitates new ways to control light. Especially, graphene is a unique material that provides gate-tunable high mobility carriers at room temperature. Graphene's linear band structure yield nearly energy independent spectrum which results in a very broad optical activity ranging from visible [14, 15] to microwave frequencies [16].

In a pioneering work by Rodrigues et al, a back-gated graphene transistor was used as a THz intensity modulator [17]. They showed that graphene can be used as an alternative low-loss THz active material. However, the back-gated transistor geometry yields limited modulation due to insufficient gating ($n<2x10^{12}$ cm$^{-2}$ and $\Delta E_F<200$ meV) [17, 18]. To overcome this limitation we have been working on electrolyte gating schemes to achieve much higher charge densities. Electrolyte gating with ionic liquids in a supercapacitor geometry produces the most efficient gating with Fermi energy shift larger than 1eV [15, 16, 19, 20]. In our previous works, we showed that graphene supercapacitors can be used to control light from visible to microwave frequencies [21]. Although, these devices show significant intensity modulation, they do not yield considerable phase modulation due to atomically thin optical path. In a recent study, passive metamaterials were integrated with gate-tunable graphene devices to achieve intensity and phase modulation in THz [22] and microwave frequencies [23]. Varying the resistance of graphene introduces electrically tunable loss to the metamaterial modulating reflected intensity and phase of THz waves.

Graphene Based Terahertz Phase Modulators

## 2. Methods and results

In this work, we show that without using any metamaterial structure, a planar graphene near a metallic surface can be used as an efficient phase modulator. Using terahertz time domain reflection spectroscopy, we studied the modulation of reflection phase from the active devices. To realize phase modulator structure, we placed gated graphene at quarter-wavelength-distance (λ/4) from a reflecting metallic surface. In the literature, this type of structure is known as Salisbury screen or the anti-reflection surface where reflecting signal is highly attenuated due to resonant absorption [24-26]. In this way, the phases of two waves, the ones reflected from graphene and metallic surface, have π difference causing destructive interference. While the reflectivity from metallic surface is constant, the reflectivity of graphene is tunable with the gate voltage [27]. Hence, by varying the conductance of graphene via ionic gating, we were able to modulate both phase and intensity of terahertz beam.

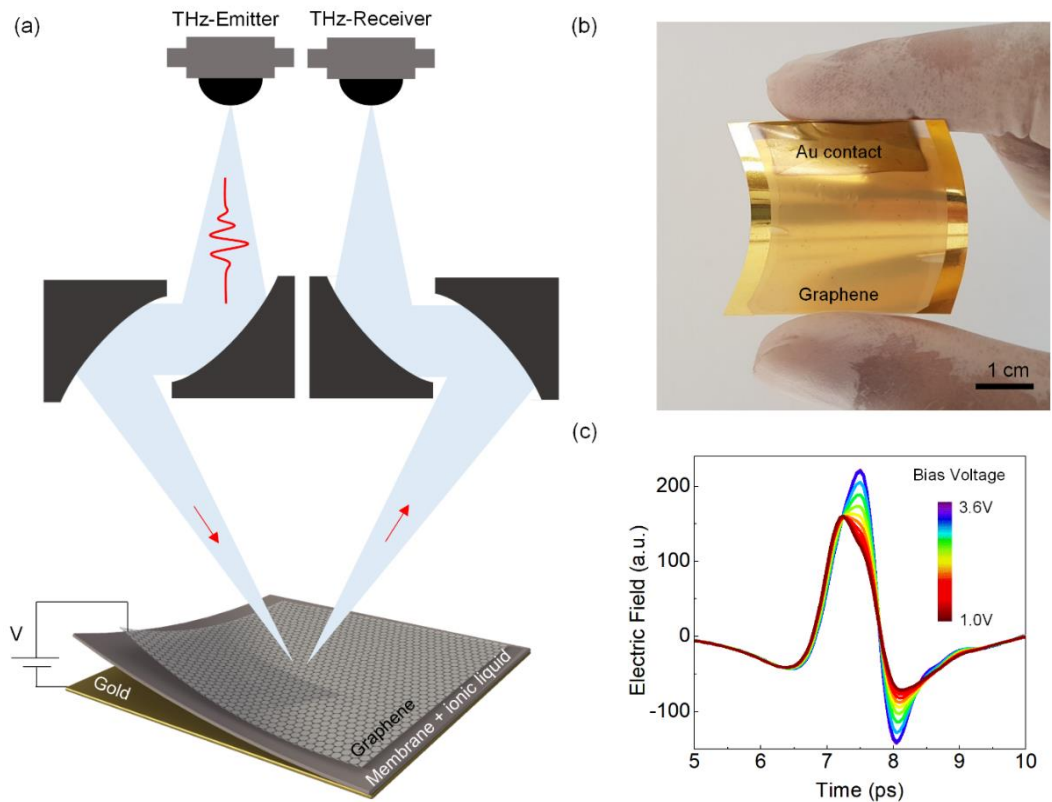

**Figure 1. Electrically tunable phase modulator.** (a) Experimental setup used for the terahertz time-domain reflection measurements. THz pulses are generated by InGaAs photoconductive antenna and focused on the device by parabolic mirrors at 10° incidence angle. The phase modulator consists of monolayer graphene on porous 20-µm-thin polyethylene membrane and gold reflector at the back. Ionic liquid electrolyte [deme][Tf2N] is injected into the membrane. (b) Photograph of the fabricated device. (c) Measured time varying electric field of terahertz pulses at gate voltages between 0 to 2V. The THz pulse is suppressed and shifted with varying applied voltage, indicating the phase modulation. Each color represents different gate voltage.

Figure 1(a) shows the schematic representation of the device layout and the THz reflection measurement system. We synthesized large area graphene on copper foils using a chemical vapor deposition method and transferred graphene on 20-µm-porous polyethylene membrane using a photoresist layer (Shipley 1813) as a mechanical support. The Raman spectrum of CVD grown graphene on $Si/SiO_2$ is illustrated in figure S1 of the supporting materials. Then, we placed graphene coated PE membrane on top of the gold reflector which operates as the gate electrode and the back reflector. The flexibility of the membrane enables us to fabricate bendable phase modulators, figure 1(b). In order to control the charge density on graphene, we soaked an electrolyte ((Diethylmethyl(2-methoxyethyl) ammonium bis(trifluoromethylsulfonyl)imide, [deme][Tf2N])) into the membrane. Graphene and the gold electrodes form a parallel plate capacitor which operates as a tunable THz cavity. By applying small gate voltage between graphene and gold electrodes, we were able to control the charge density ($\sim 10^{14}$ $cm^{-2}$) and Fermi energy ($> 1$ eV) of graphene. Figure S2 in the supporting information shows the Fermi energy shift of graphene with respect to applied voltage. We measured the complex THz reflectivity from the device using a time domain terahertz system (Toptica Teraflash System). The system has two fiber-coupled InGaAs antennas which can generate and detect terahertz pulses with >5 THz bandwidth and over 90 dB dynamic range. Using four 1" parabolic mirrors (Toptica THz reflection head), the terahertz pulse is focused on the sample at 10° incidence angle. Figure 1(c) shows the measured time varying electric field of terahertz pulse at gate voltages from 0 to 2 V. As the gate voltage changes, the reflected THz pulse shows a significant temporal shift with diminishing shape, which also indicates the modulation of the phase and intensity [12, 28, 29].

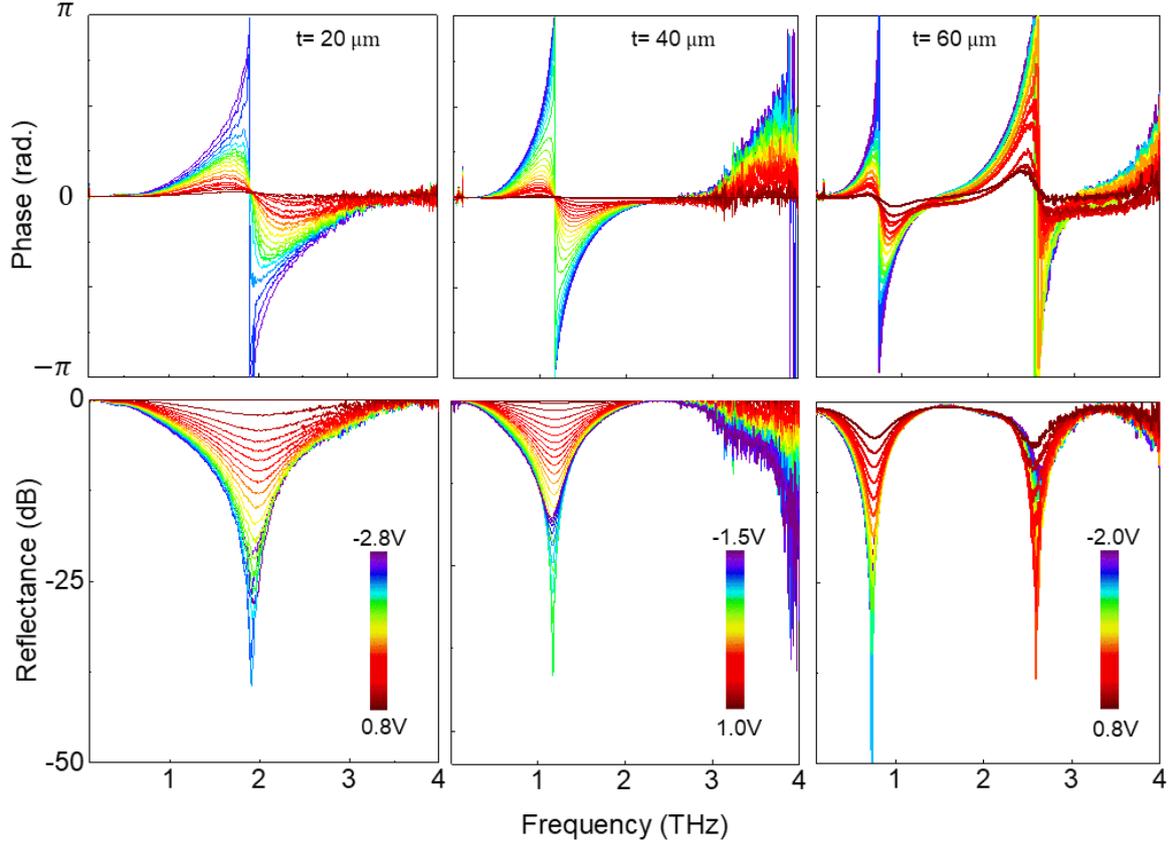

**Figure 2. Phase and amplitude modulation.** Variation of phase (radians) and intensity (dB) of the THz reflection of the devices with 20, 40 and 60 $\mu m$ cavity lengths. The thickness of the membrane defines the cavity length and the resonance frequencies. When the critical coupling condition achieved, the reflectivity is suppressed and the phase shows a step like change around the resonance.

Fourier transforming the time varying electric field provides the phase and amplitude spectrum. We observed clear multiple resonances in the THz reflection spectrum. The resonance frequency depends on the thickness of the membrane which defines the cavity length, $l = \lambda(2m + 1)/4n$ where, $n$ is the index of refraction, $\lambda$ is the resonance wavelength, and $m$ is an integer number. Around the resonance, the phase undergoes a $\pi$-shift and the reflection intensity shows a suppression of nearly 50 dB. Figure 2 demonstrates the modulation of phase (relative to the phase at Dirac point) and the normalized reflectance spectrum for three devices with cavity lengths 20, 40, and 60 µm. The color bar indicates the gate voltages varying from charge neutrality point to the highly doped case of graphene. The charge neutrality point of graphene occurs at nearly 1 V when it makes contact with ionic liquid medium and gold electrode (figure S3). We observed a clear phase and intensity modulation over a broad spectral range. As the membrane gets thicker, the resonance shifts to longer wavelengths due to longer cavity length. We also observed a slight shift in the resonance frequency with the increasing gate voltage.

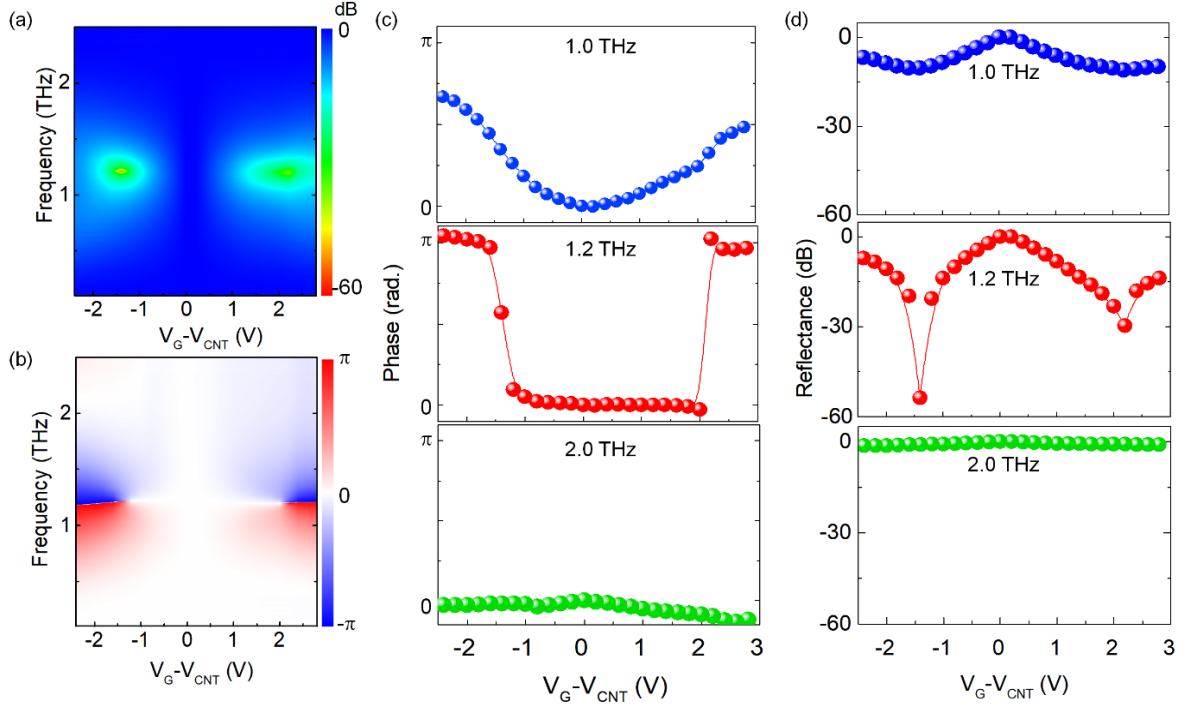

**Figure 3. Voltage controlled phase.** Two-dimensional maps showing the variation of the reflection amplitude (a) and the reflection phase (b) as a function of the frequency and the gate voltage. The color bar indicated the values of reflection (dB) and phase (radian). (c) and (d) show the variation of phase and amplitude with the gate voltage for 1.0, 1.2, and 2.0 THz frequencies.

Our device provides efficient means of controlling both intensity and phase of THz waves. Figures 3(a) and 3(b) show the two-dimensional maps of the THz reflectance and phase as a function of frequency and applied voltage for the 40 µm cavity length device, respectively. The color scales constitute to the intensity from 0 to 60 dB and phase from $\pi$ to $-\pi$. Two resonance dips occurred at 1.2 THz for negative and positive voltages. The sharp area is indicated in red color for the reflectance values of 50 dB. Nearly perfect electron-hole symmetry in graphene resulted in two resonances for both hole and electron doping. Because of the causal relation the reflection phase and amplitude varies with the gate voltage. The line profile of the phase and reflectance is represented in figures 3(c) and 3(d) respectively, at frequencies of 1.2, 1.0 and 2.0 THz. These frequencies represent three different behaviors; resonance, near-resonance and off-resonance. At the resonance frequency (1.2 THz), the graph shows the π-phase shift which occurred at positive and negative gate voltages. At near-resonance frequencies (1.0 THz), the reflection phase gradually increases to $\pi/2$ at highly doped case. At off-resonance frequencies (2 THz), the phase is nearly constant. Using transfer matrix method we simulated the resonance and off-resonance frequencies with respect to surface impedance. The resultant figures are demonstrated in the supporting materials (Figure S4). Similarly, in figure 3(d), the intensity variations are depicted for mentioned three cases yielding intensity modulation of 55, 30, 10 dB respectively.

To understand more insight about the operation of the device, we would like to develop a quantitative model using the concept of wave impedance $Z = \frac{E}{H}$ where E and H denotes the electric and magnetic field of the wave along the surface. In our device geometry, graphene operates as a tunable impedance surface at THz frequencies. Electrostatic control of charge density on graphene yield tunable optical conductivity, σ (ω) which defines the surface impedance of graphene as $Z_g = \frac{Z_0}{1+Z_0\sigma(\omega)}$ where $Z_0=377\Omega$ is the free space impedance and σ(ω) is the frequency dependent optical conductivity of graphene. We can describe the tunable optical conductance of graphene with Drude model as,

$$\sigma(\omega) = \frac{\sigma_{DC}}{1 - i\omega\tau}$$

where $\sigma_{DC}$ is the DC conductivity of graphene, $\omega$ is angular frequency, and $\tau$ is carrier scattering time. Here, the sheet conductivity of graphene, $\sigma_{DC}$ relies on the Fermi energy $\sigma_{DC} = \left(\frac{e^2}{\pi\hbar^2}\right) E_F \tau$ [30]. For microwave frequencies ($\omega \ll \tau^{-1}$), the optical conductivity equals to $\sigma_{DC}$ and the surface impedance of graphene has only real values. However for THz frequencies, the scattering rate is comparable with the excitation frequency ($\omega \sim \tau^{-1}$), hence, resulting in a complex surface impedance. Note that both $\sigma_{DC}$ and $\tau$ varies with charge concentration.

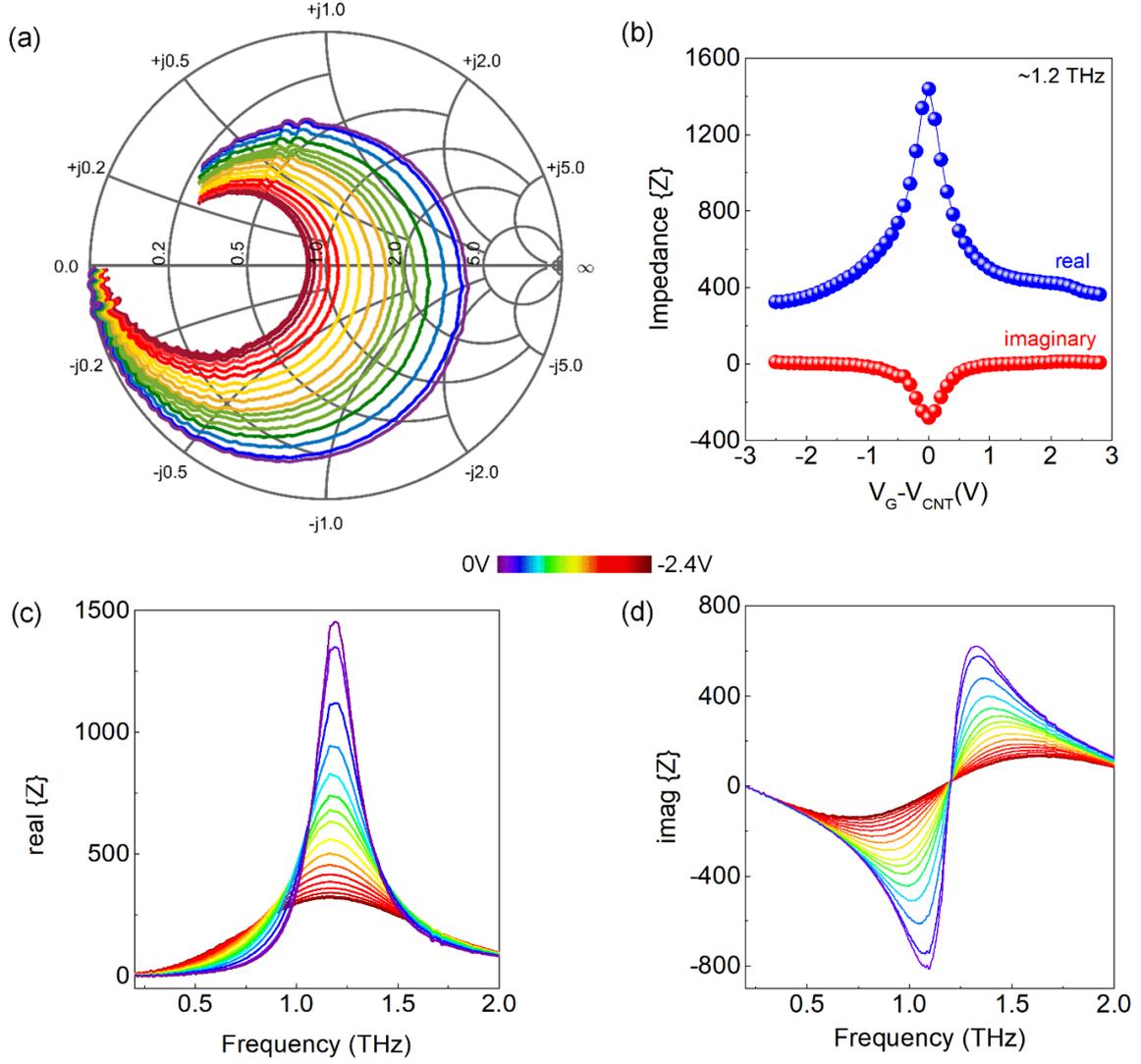

**Figure 4. Tunable impedance surface.** (a) Smith chart representation of the complex THz reflectivity which allows us to extract surface impedance of the device. (b) The variation of both the real (x-axis on the chart) and imaginary parts of the impedance at resonance extracted from Smith chart as a function of positive and negative gate voltages. The real (c) and imaginary (d) parts of the complex impedance plotted with respect to frequency.

We can extract the surface impedance of the device from time domain terahertz spectroscopy which provides complex Fresnel coefficients. Plotting complex reflectivity in polar coordinates (Smith chart representation, figure 4(a) provides wealth of information about the variation of the impedance of the device. The Fresnel coefficient of reflection of the device can be written as $\tilde{r} = \frac{Z-Z_0}{Z+Z_0}$. Using this equation, the surface impedance of the device can be extracted from the Smith chart as $Z = Z_0 \frac{1+\tilde{r}}{1-\tilde{r}}$. Figures 4(b) and 4(c) show the extracted real and imaginary parts of the surface impedance at different gate voltages. At the Dirac point, the resistive part of surface impedance shows a peak which reaches to a maxima at the resonance frequency. As the gate voltage increases the surface impedance decreases and its spectral distribution gets broader. The reactive part of the surface impedance shows undulation from negative to positive values. Around resonance frequencies the imaginary impedance is nearly at zero crossing reaching extremum at Dirac point. The variation of the real and imaginary impedance are plotted in figure 4(c). The real part of the impedance corresponds to the resistance of graphene which changed from 325 Ω to 1450 Ω with respect to the applied voltage. Unlike microwave frequencies, for THz, the imaginary part of the impedance is nonzero effecting the resonance conditions.

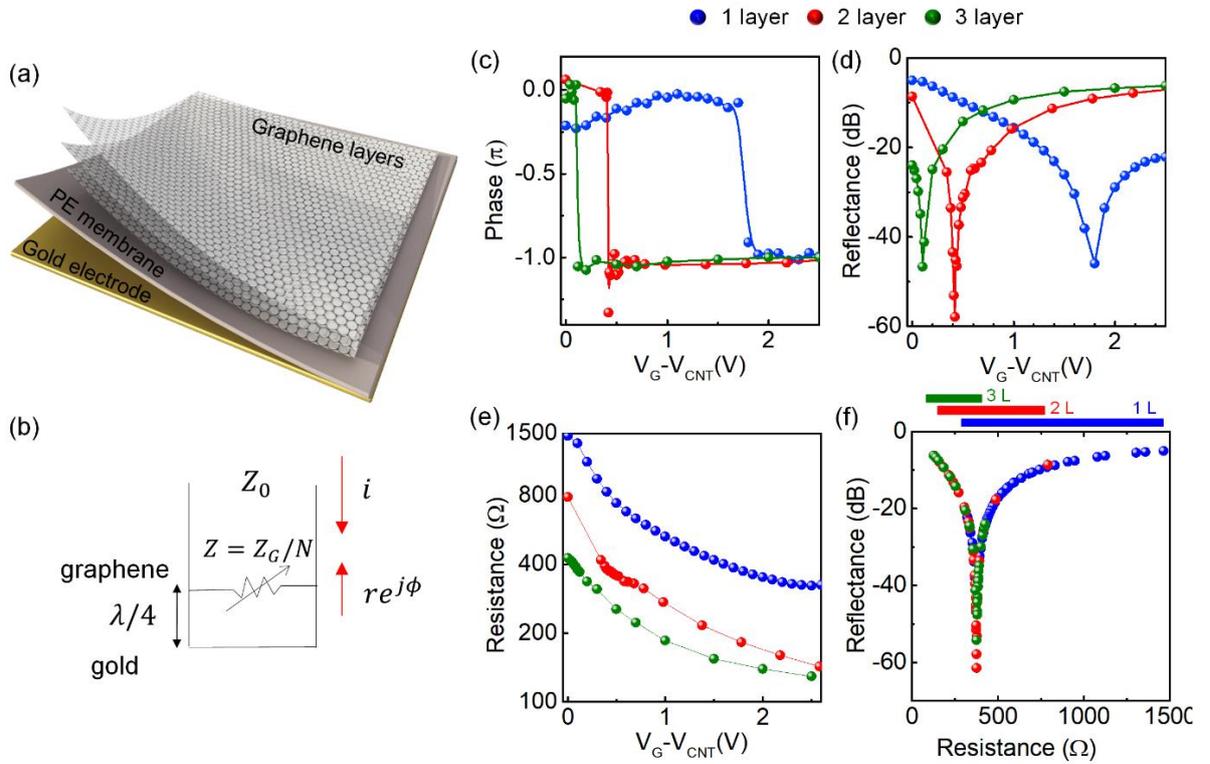

**Figure 5. Multilayer phase modulator.** (a) Schematic representation of the THz phase modulator using few-layer graphene electrode. (b) Small signal model of the device. Graphene is represented with a variable resistor. The membrane is modeled as a quarter-wave transmission line and the gold reflector is represented with a short circuit at the end of the transmission line. (c) Phase and (d) reflectance modulation of single, two, and three layers of graphene at 1 THz frequency. As the layer number increases, the critical coupling condition ($Z_D = Z_0 = 377\ \Omega$) is achieved at lower gate voltages. (e) Extracted sheet resistance of single, two, and three layers of graphene as a function of applied voltage. The minimum resistance for single, bilayer and trilayer graphene are 325, 120 and 100 $\Omega$ respectively. (f) Reflectance versus resistance curves for single, two, and three layers of graphene.

The modulation bandwidth is directly related to the tunability of the resistance of graphene which varies between 0.3 to 1.5 k$\Omega$ for single layer. This tunability window is limited by the unintentional doping around Dirac point, and the electrochemical stability of the ionic liquid at high gate voltages. Increasing the number of graphene layers could extend the range of tunability to lower resistance. To test this idea, we fabricated devices with few-layer graphene which were obtained by sequential transfer process of single layer graphene. Figure 5(a) shows the schematic representation of the THz phase modulator with few-layer graphene on 40 µm thick PE membrane. The blue, red, and green colors correspond to single, two, and tri-layer graphene devices, respectively. The small signal model of the device is shown in figure 5(b). The effective surface impedance of the device $Z=Z_G/N$ decreases with the layer number, $N$. Figures 5(c) and 5(e) compares the variation of THz reflection phase and intensity at 1.2 THz for single, bilayer and three layer graphene devices. The device with single layer graphene electrode shows the resonance absorption at 1.8 V gate voltage, however, the device with three layer graphene electrode shows the resonance at much lower gate voltage of 0.1 V due to lower resistance. From the reflectivity spectra, we extracted variation of the resistance of single, two, and three layers graphene with the gate voltage, figure 5(d). At the Dirac point, the maximum resistance is 1500, 800 and 400 $\Omega$ respectively. Our experimental results show that graphene layers operate as isolated single layers. We observed linear scaling of the sheet resistance with the number of graphene layer. However, similar devices with bilayer graphene synthesized on copper would provide interesting optical response in THz wavelengths due to small electronic band gap. Similarly, at high doping, the minimum resistance for single, bilayer and trilayer graphene are 325, 120 and 100 $\Omega$ respectively. Since the resonance condition is governed by the surface impedance of the graphene layer, devices with different layer number show the same scaling behaviors with resistance, figure 5(f). For all devices, the resonance condition is achieved at nearly $Z=377$ $\Omega$. The operational ranges of single, two, and three layer graphene devices are indicated with respective color, figure 5(f).

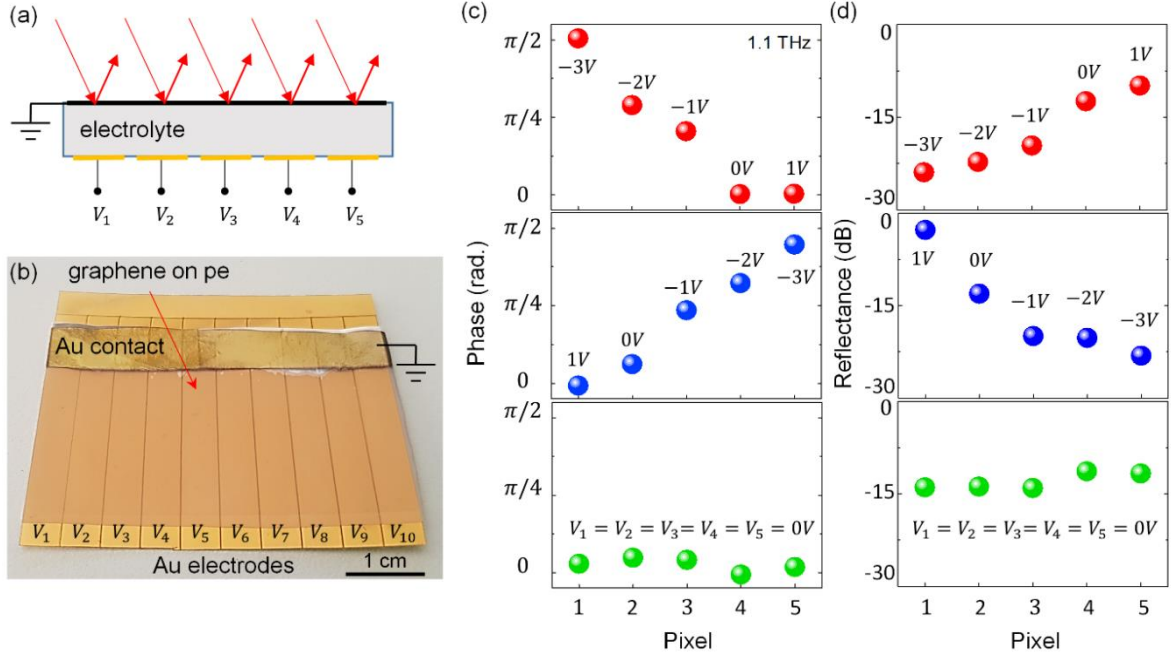

**Figure 6. Electrically tunable phased-array.** (a) Schematic side view of the pixelated phased array which consists of continuous graphene electrodes and pattered gate-electrodes. Individually addressable gold electrodes allows local control of the charge density and the reflection phase. (b) Photograph of fabricated device. (c) Spatial variation of the reflection phase (at 1.1 THz) measured from the multipixel device under different voltage configuration. (d) Corresponding reflectance variations of THz waves at 1.1 THz.

We anticipate that the demonstrated phase modulation could enable new type of THz devices. To show the promises of our approach, we would like to demonstrate a multipixel phased array. Figure 6(a) shows the schematic side view of the phased array. We used continuous 3x5 cm² single layer graphene on PE as the top layer. Then we pattered the back electrode (80 nm evaporated gold on flexible substrate) into to 1-dimensional arrays of ribbons. Figure 5(b) shows the photograph of the fabricated device. We grounded the continuous graphene and applied spatially changing gate voltages to each column. Ability to control the local charge density with the pattered electrodes enable us to generate a phase gradient on the surface. Figure 6(c) shows the measured reflection phase from the pixels at three different voltage configurations for 1.1 THz. In the first case, we applied a gradient voltage to generate a linear spatial phase variation from $\pi/2$ to 0 radians. In a similar way, when we reversed the gate voltage by applying consecutive 1 V to -3 V from first to last columns, the phase modulation changed from 0 to nearly $\pi/2$ respectively. In the last configuration, we kept the gate voltage at zero for all

Graphene Based Terahertz Phase Modulators

pixels, and the relative phase change was nearly zero radian at 1.1 THz. Figure 6(d) shows the corresponding reflectance variations measured from multipixel device with respect to applied voltages. Using photolithography, the pixel size can be miniaturized while extending the density of THz phased-array. A thin film structure can be regarded as a metasurface if the thickness is much smaller than the wavelength [31]. However in our case, the thickness of the device is quarter of the wavelength which is not suitable to classify our structure as a metasurface. In the pixelated device, the lateral dimensions are much longer than the wavelength. It is more suitable to consider these surfaces as phased arrays. These devices can be regarded as a metadevice when the pixels size is much smaller than the wavelength. We anticipate that scaling the lateral dimension of the pixels to sub wavelength scale would provide new possibility to control not only phase but also polarization state of reflected light.

## 3. Conclusion

As a conclusion, we demonstrated an efficient THz phase modulator using active graphene devices. Our device comprises electrostatically gated graphene layer placed at quarter-wave distance from a metallic reflector. In this configuration, graphene behaves as a tunable impedance surface modulating reflection phase. At resonance, the phase of terahertz wave undergoes step-like $\pi$ change while the intensity of the wave is highly attenuated (~50 dB). Efficient carrier density generation at graphene surface is a key parameter that enabled us to modulate the phase of THz radiation. We were also able to extract the complex impedance of electrically gated graphene for THz frequencies. The practicality of device fabrications allowed us to design a tunable phased array on a flexible substrate which operates as a gradient phase surface.

**Acknowledgment**: This work was supported by the European Research Council (ERC) Consolidator Grant No. ERC-682723 SmartGraphene. C.K. acknowledges BAGEP Award of the Science Academy.

**References:**

[1] Shields A J 2007 Semiconductor quantum light sources *Nat Photonics* **1** 215-23
[2] Nowak A K, Portalupi S L, Giesz V, Gazzano O, Dal Savio C, Braun P F, Karrai K, Arnold C, Lanco L, Sagnes I, Lemaitre A and Senellart P 2014 Deterministic and electrically tunable bright single-photon source *Nat Commun* **5**

[3]     Unlu M, Hashemi M R, Berry C W, Li S, Yang S H and Jarrahi M 2014 Switchable Scattering Meta-Surfaces for Broadband Terahertz Modulation *Sci Rep-Uk* **4**
[4]     Shi S F, Zeng B, Han H L, Hong X, Tsai H Z, Jung H S, Zettl A, Crommie M F and Wang F 2015 Optimizing Broadband Terahertz Modulation with Hybrid Graphene/Metasurface Structures *Nano Lett* **15** 372-7
[5]     Chen H T, Padilla W J, Zide J M O, Gossard A C, Taylor A J and Averitt R D 2006 Active terahertz metamaterial devices *Nature* **444** 597-600
[6]     Chen H T, O'Hara J F, Taylor A J, Averitt R D, Highstrete C, Lee M and Padilla W J 2007 Complementary planar terahertz metamaterials *Opt Express* **15** 1084-95
[7]     Kersting R, Strasser G and Unterrainer K 2000 Terahertz phase modulator *Electron Lett* **36** 1156-8
[8]     Chen C Y, Hsieh C F, Lin Y F, Pan R P and Pan C L 2004 Magnetically tunable room-temperature 2 pi liquid crystal terahertz phase shifter *Opt Express* **12** 2625-30
[9]     Lee S H, Choi M, Kim T T, Lee S, Liu M, Yin X, Choi H K, Lee S S, Choi C G, Choi S Y, Zhang X and Min B 2012 Switching terahertz waves with gate-controlled active graphene metamaterials *Nat Mater* **11** 936-41
[10]    Harrison P, Kelsall R W, Kinsler P and Donovan K 1996 Quantum well intersubband transitions as a source of terahertz radiation *The Ninety Eight - 1998 Ieee Sixth International Conference on Terahertz Electronics Proceedings*  74-8
[11]    Borenstain S I and Katz J 1989 Evaluation of the Feasibility of a Far-Infrared Laser Based on Intersubband Transitions in Gaas Quantum Wells *Appl Phys Lett* **55** 654-6
[12]    Chen H T, Padilla W J, Cich M J, Azad A K, Averitt R D and Taylor A J 2009 A metamaterial solid-state terahertz phase modulator *Nat Photonics* **3** 148-51
[13]    Watts C M, Shrekenhamer D, Montoya J, Lipworth G, Hunt J, Sleasman T, Krishna S, Smith D R and Padilla W J 2014 Terahertz compressive imaging with metamaterial spatial light modulators *Nat Photonics* **8** 605-9
[14]    Nair R R, Blake P, Grigorenko A N, Novoselov K S, Booth T J, Stauber T, Peres N M R and Geim A K 2008 Fine structure constant defines visual transparency of graphene *Science* **320** 1308-
[15]    Polat E O and Kocabas C 2013 Broadband Optical Modulators Based on Graphene Supercapacitors *Nano Lett* **13** 5851-7
[16]    Balci O, Polat E O, Kakenov N and Kocabas C 2015 Graphene-enabled electrically switchable radar-absorbing surfaces *Nat Commun* **6**
[17]    Sensale-Rodriguez B, Yan R S, Kelly M M, Fang T, Tahy K, Hwang W S, Jena D, Liu L and Xing H G 2012 Broadband graphene terahertz modulators enabled by intraband transitions *Nat Commun* **3**
[18]    Ren L, Zhang Q, Yao J, Sun Z Z, Kaneko R, Yan Z, Nanot S, Jin Z, Kawayama I, Tonouchi M, Tour J M and Kono J 2012 Terahertz and Infrared Spectroscopy of Gated Large-Area Graphene *Nano Lett* **12** 3711-5
[19]    Kakenov N, Takan T, Ozkan V A, Balci O, Polat E O, Altan H and Kocabas C 2015 Graphene-enabled electrically controlled terahertz spatial light modulators *Opt Lett* **40** 1984-7
[20]    Kakenov N, Balci O, Polat E O, Altan H and Kocabas C 2015 Broadband terahertz modulators using self-gated graphene capacitors *J Opt Soc Am B* **32** 1861-6
[21]    Kakenov N, Balci O, Takan T, Ozkan V A, Akan H and Kocabas C 2016 Observation of Gate-Tunable Coherent Perfect Absorption of Terahertz Radiation in Graphene *Acs Photonics* **3** 1531-5
[22]    Miao Z Q, Wu Q, Li X, He Q, Ding K, An Z H, Zhang Y B and Zhou L 2015 Widely Tunable Terahertz Phase Modulation with Gate-Controlled Graphene Metasurfaces *Phys Rev X* **5**
[23]    Balci O, Karademir E, Cakmakyapan S, Kakenov N, Ozan Polat E, Caglayan H, Ozbay E and Kocabas C 2015 Electrically Switchable Metadevices. In: *ArXiv e-prints*,
[24]    Tennant A and Chambers B 2004 Adaptive radar absorbing structure with PIN diode controlled active frequency selective surface *Smart Mater Struct* **13** 122-5
[25]    Woo J M, Kim M S, Kim H W and Jang J H 2014 Graphene based salisbury screen for terahertz absorber *Appl Phys Lett* **104**

[26] Jang M S, Brar V W, Sherrott M C, Lopez J J, Kim L, Kim S, Choi M and Atwater H A 2014 Tunable large resonant absorption in a midinfrared graphene Salisbury screen *Phys Rev B* **90**

[27] Sensale-Rodriguez B, Yan R S, Rafique S, Zhu M D, Li W, Liang X L, Gundlach D, Protasenko V, Kelly M M, Jena D, Liu L and Xing H G 2012 Extraordinary Control of Terahertz Beam Reflectance in Graphene Electro-absorption Modulators *Nano Lett* **12** 4518-22

[28] Hsieh C F, Pan R P, Tang T T, Chen H L and Pan C L 2006 Voltage-controlled liquid-crystal terahertz phase shifter and quarter-wave plate *Opt Lett* **31** 1112-4

[29] Lin X W, Wu J B, Hu W, Zheng Z G, Wu Z J, Zhu G, Xu F, Jin B B and Lu Y Q 2011 Self-polarizing terahertz liquid crystal phase shifter *Aip Adv* **1**

[30] Liu F L, Chong Y D, Adam S and Polini M 2014 Gate-tunable coherent perfect absorption of terahertz radiation in graphene *2d Mater* **1**

[31] Yu N and Capasso F 2014 Flat optics with designer metasurfaces *Nature materials* **13** 139-50

**Supporting materials for "Graphene Based Terahertz Phase Modulators"**

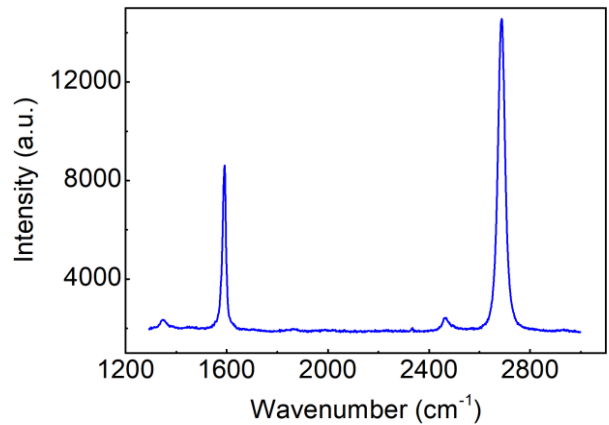

**Figure S1**. Raman spectrum of CVD grown graphene on Si/SiO$_2$ substrate. The intensity ratio of 2D/G is 1.7.

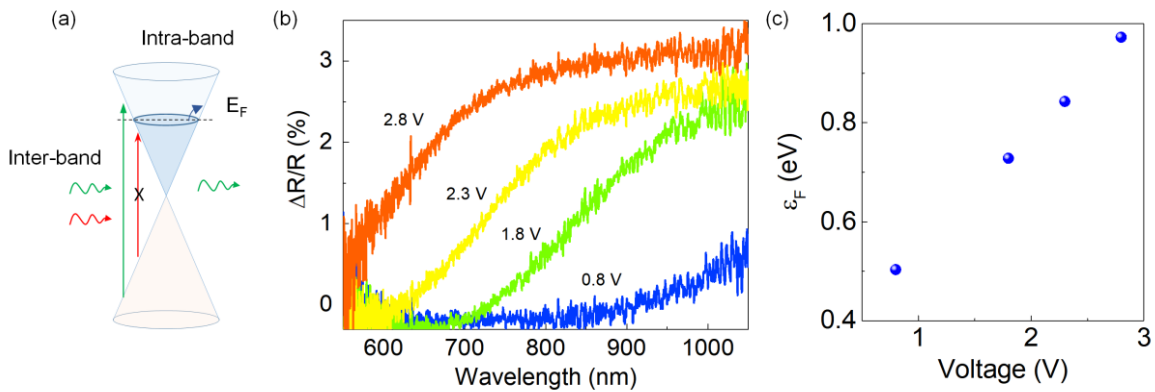

**Figure S2**. (a) Inter-band absorption mechanism of graphene. By shifting the Fermi energy of graphene to higher levels, more energetic photons are blocked from absorption causing increase in the transmission or reflection. (b) Reflection spectrum of electrostatically gated graphene. (c) Extracted Fermi energy values for corresponding applied voltage.

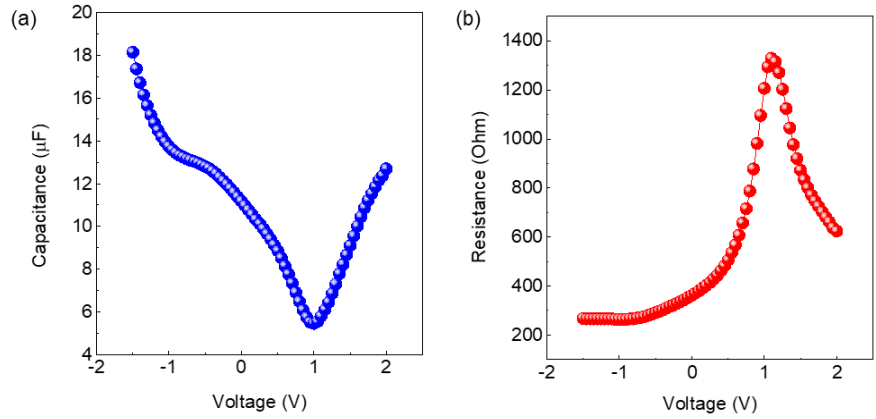

**Figure S3.** LCR meter measurement of graphene ionic liquid gold metal structure of capacitance (a) and resistance (b) as a function of applied voltage sweep (Keysight (Agilent/HP) 4284A). The maximum value of resistance curve (b) indicates the Dirac point of graphene (1.1 V). LCR meter cannot measure accurately the low resistance values; hence, other methods must be applied to measure the sheet resistance of highly doped graphene.

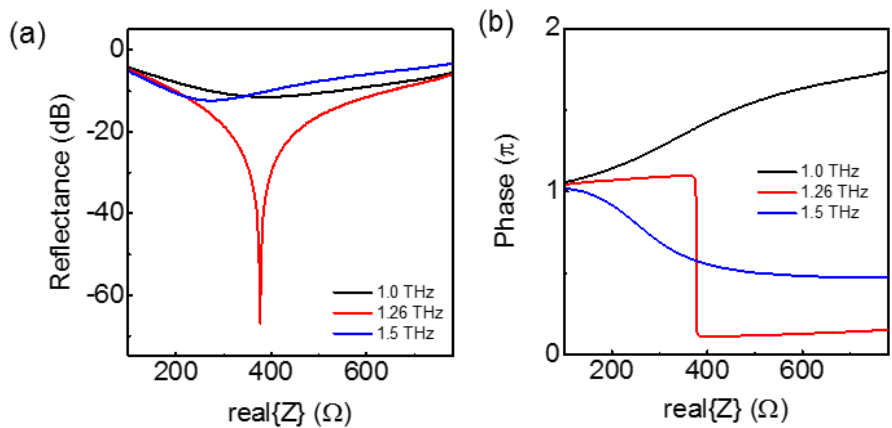

**Figure S4.** Controlling phase without intensity modulation. Variation of the reflection (a) and phase (b) from the device as a function of impedance of the graphene layer. When the impedance matches the free space impedance, the reflection is at minimum value with $\pi$ shift in phase of the THz wave.